# Equivalence of the empirical shifted Deng-Fan oscillator potential for diatomic molecules

M. Hamzavi[1*], S. M. Ikhdair[2 +], K.-E. Thylwe[3 †]

[1]*Department of Basic Sciences, Shahrood Branch, Islamic Azad University, Shahrood, Iran*
[2]*Physics Department, Near East University, 922022 Nicosia, North Cyprus, Mersin 10, Turkey*
[3]*KTH-Mechanics, Royal Institute of Technology, S-100 44 Stockholm, Sweden*
[*]*Corresponding author: Tel.:+98 273 3395270, fax: +98 273 3395270*
[*]Email: majid.hamzavi@gmail.com
Email: sikhdair@neu.edu.tr; sikhdair@gmail.com [+]
[†] Email: ket@mech.kth.se

**Abstract**

**W**e obtain the bound-state solutions of the radial Schrödinger equation (SE) with the shifted Deng-Fan (sDF) oscillator potential in the frame of the Nikiforov-Uvarov (NU) method and employing Pekeris-type approximation to deal with the centrifugal term. The analytical expressions for the energy eigenvalues and the corresponding wave functions are obtained in closed form for arbitrary $l$-state. The ro-vibrational energy levels for a few diatomic molecules are also calculated. They are found to be in good agreement with those ones previously obtained by the Morse potential.

## 1. Introduction

The most interesting phenomenon in many fields of physics and chemistry is to obtain the exact analytical solutions of the fundamental wave equations. These exact solutions play an important role in quantum mechanics since they contain all the necessary information regarding the quantum system under study. The two typical examples in quantum mechanics are the exact solution of the SE for a hydrogen atom (Coulomb field) and the harmonic oscillator [1-3]. The Mie-type and pseudo-harmonic oscillator potentials are also two exactly solvable potentials [4-5]. Also, there are much potentials that are exactly solvable for $s$-wave (i.e. $l=0$). However, their analytical exact solutions cannot be determined for $l \neq 0$ and hence approximation methods are used to solve them [6-23].

The choice of potential is crucial for molecular spectroscopy and molecular dynamics. The ideal potential is chosen to behave properly at its limits of coordinates, i.e., $V(0) = \infty$ and $V(\infty)$ approaches to constant. For stable molecule the potential has a minimum at the equilibrium bond length $r_e$, i.e., $V(r_e) = 0$ and $V''(r_e) > 0$.

In 1929, Morse proposed the first most widely used three-parameter solvable empirical potential energy function for diatomic molecules having the form $V(r - r_e) = V(q) = D\left[1 - e^{-\alpha q}\right]^2$, in which $D$ is the dissociation energy, $r_e$ is the equilibrium bond length and $\alpha$ denotes the range of the potential. It is simplest practical anharmonic oscillator model which allows for dissociation but the bond length approaches 0 for which the Morse potential goes to a large value whereas the actual potential should approach infinity. This inherent shortcoming of the Morse potential leads to a small wave function but not 0 for bound vibrational states as the bond length approaches 0.

In 1957, Deng and Fan [24] proposed a simple potential model for diatomic molecules called the Deng-Fan (DF) oscillator potential. The DF potential was called a general Morse potential [25] whose analytical expressions for energy levels and wave functions have been derived [24,25] and related to the Manning-Rosen potential [26] (also called Eckart potential by some authors [27]) is anharmonic potential. It has the correct physical boundary conditions at $r = 0$ and $\infty$, and is defined by

$$V(r) = D\left[1 - \frac{b}{e^{\alpha r} - 1}\right]^2, \qquad b = e^{\alpha r_e} - 1, \ r \in (0, \infty) \tag{1a}$$

and also the shifted DF (sDF) potential is

$$V(r) = D\left[1 - \frac{b}{e^{\alpha r} - 1}\right]^2 - D = D\left[\frac{b^2}{\left(e^{\alpha r} - 1\right)^2} - \frac{2b}{e^{\alpha r} - 1}\right], \ r \in (0, \infty) \tag{1b}$$

where the three positive parameters $D$, $r_e$ and $\alpha$ stand for the dissociation energy, the equilibrium inter-nuclear distance and the range of the potential well, respectively.

The DF potential (1a) is qualitatively similar to the Morse potential but has the correct asymptotic behavior as the inter-nuclear distance approaches 0 [25] and used to describe diatomic molecular energy spectra and electromagnetic transitions.

The Morse and the sDF potentials are very close to each other for large values of in the regions $r \sim r_e$ and $r > r_e$, but are very different at $r \sim 0$. Further, if the two potentials are deep ($D \gg 1$), they could be well approximated by a harmonic oscillator in the region $r \sim r_e$ [25]. In Fig. 1, we plot both the sDF potential and the Morse potential using the parameters set for $H_2$ diatomic molecule given by $D = 4.74441001\, eV$, $\alpha = 1.9426\, A^{\circ -1}$ and $r_e = 0.7416\, A^{\circ}$.

When the parameter $\alpha$ in Eq. (1a) goes to zero, the limits of the DF oscillator potential becomes the Kratzer-Fues molecular potential [28,29], namely, $\lim_{\alpha \to 0} V_{DF}(r) = D\left(\dfrac{r-r_e}{r}\right)^2$. The Kratzer-Fues molecular potential has been extensively used for investigating the properties of diatomic molecules [30-32].

Mesa *et al.* studied the exact solvability of the $s$-wave bound state energy eigenvalues and eigenfunctions of the generalized Morse potential due to the fact that it belongs to the class of the Eckart potential, a member of the hypergeometric Natanzon potentials which can be solved algebraically by means of SO(2,2) symmetry algebra [25]. Dong and Gu have approximately presented the bound state solutions of the SE with the Deng-Fan oscillator interaction [33]. Ikhdair solved the Dirac equation for the Deng-Fan oscillator by using an improved approximation scheme to deal with the centrifugal term [34].

The aim of this work is to obtain the approximate bound state energy eigenvalue equation and the corresponding normalized wave functions for the diatomic molecules subject to the empirical sDF oscillator interaction using the parametric NU method [34-37]. We also investigate the equivalence of the sDF oscillator potential and compare it with the Morse potential according to quantitative tests on four molecules.

The present analytical solution to the Schrödinger equation (SE) associated with sDF potential is useful since the eigenvalues and wave functions permit exact determination of transition frequencies, matrix elements and oscillator strengths. Furthermore, we apply our results to obtain the numerical ro-vibrational spectrum of some diatomic molecules and compare with those ones obtained with the Morse potential.

The paper is organized as follows. In Section 2, we present the parametric NU method. In Section 3, we solve the radial SE for the empirical sDF oscillator potential to obtain the energy spectrum and the corresponding wave functions. The energy levels for a few diatomic molecules are also presented. Finally, the conclusion is given in Section 4.

**2. Parametric NU method**

The NU method is used to solve second order differential equations with an appropriate coordinate transformation $s = s(r)$ [35]

$$\psi_n''(s) + \frac{\tilde{\tau}(s)}{\sigma(s)}\psi_n'(s) + \frac{\tilde{\sigma}(s)}{\sigma^2(s)}\psi_n(s) = 0, \tag{2}$$

where $\sigma(s)$ and $\tilde{\sigma}(s)$ are polynomials, at most of second degree, and $\tilde{\tau}(s)$ is a first-degree polynomial. To make the application of the NU method simpler and direct without need to check the validity of solution. We present a shortcut for the method. So, at first we write the general form of the Schrödinger-like equation (2) in a more general form applicable to any potential as follows [36,37]

$$\psi_n''(s) + \left(\frac{c_1 - c_2 s}{s(1 - c_3 s)}\right)\psi_n'(s) + \left(\frac{-As^2 + Bs - C}{s^2(1 - c_3 s)^2}\right)\psi_n(s) = 0, \tag{3}$$

satisfying the wave functions

$$\psi_n(s) = \phi(s) y_n(s). \tag{4}$$

Comparing (3) with its counterpart (2), we obtain the following identifications:

$$\tilde{\tau}(s) = c_1 - c_2 s, \quad \sigma(s) = s(1 - c_3 s), \quad \tilde{\sigma}(s) = -As^2 + Bs - C. \tag{5}$$

Following the NU method [35], we obtain the followings [36],

(i) the relevant constant:

$$c_4 = \frac{1}{2}(1 - c_1), \qquad c_5 = \frac{1}{2}(c_2 - 2c_3),$$

$$c_6 = c_5^2 + A, \qquad c_7 = 2c_4 c_5 - B,$$

$$c_8 = c_4^2 + C, \qquad c_9 = c_3(c_7 + c_3 c_8) + c_6,$$

$$c_{10} = c_1 + 2c_4 + 2\sqrt{c_8} - 1 > -1, \qquad c_{11} = 1 - c_1 - 2c_4 + \frac{2}{c_3}\sqrt{c_9} > -1, \ c_3 \neq 0,$$

$$c_{12} = c_4 + \sqrt{c_8} > 0, \qquad c_{13} = -c_4 + \frac{1}{c_3}(\sqrt{c_9} - c_5) > 0, \ c_3 \neq 0. \tag{6}$$

(ii) the essential polynomial functions:

$$\pi(s) = c_4 + c_5 s - \left[\left(\sqrt{c_9} + c_3\sqrt{c_8}\right)s - \sqrt{c_8}\right], \tag{7}$$

$$k = -(c_7 + 2c_3 c_8) - 2\sqrt{c_8 c_9}, \tag{8}$$

$$\tau(s) = c_1 + 2c_4 - (c_2 - 2c_5)s - 2\left[\left(\sqrt{c_9} + c_3\sqrt{c_8}\right)s - \sqrt{c_8}\right], \tag{9}$$

$$\tau'(s) = -2c_3 - 2\left(\sqrt{c_9} + c_3\sqrt{c_8}\right) < 0. \tag{10}$$

(iii) The energy equation:

$$c_2 n - (2n+1)c_5 + (2n+1)\left(\sqrt{c_9} + c_3\sqrt{c_8}\right) + n(n-1)c_3 + c_7 + 2c_3 c_8 + 2\sqrt{c_8 c_9} = 0. \tag{11}$$

(iv) The wave functions

$$\rho(s) = s^{c_{10}}(1 - c_3 s)^{c_{11}}, \tag{12}$$

$$\phi(s) = s^{c_{12}}(1 - c_3 s)^{c_{13}}, \ c_{12} > 0, \ c_{13} > 0, \tag{13}$$

$$y_n(s) = P_n^{(c_{10}, c_{11})}(1 - 2c_3 s), \ c_{10} > -1, \ c_{11} > -1, \tag{14}$$

$$\psi_{nl}(s) = N_{nl} s^{c_{12}}(1 - c_3 s)^{c_{13}} P_n^{(c_{10}, c_{11})}(1 - 2c_3 s). \tag{15}$$

where $P_n^{(\mu,\nu)}(x)$, $\mu > -1$, $\nu > -1$, and $x \in [-1, 1]$ are Jacobi polynomials with

$$P_n^{(\alpha,\beta)}(1 - 2s) = \frac{(\alpha+1)_n}{n!} {}_2F_1(-n, 1 + \alpha + \beta + n; \alpha + 1; s), \tag{16}$$

and $N_{nl}$ is a normalization constant. Also, the above wave functions can be expressed in terms of the hypergeometric function as

$$\psi_{nl}(s) = N_{nl} s^{c_{12}}(1 - c_3 s)^{c_{13}} {}_2F_1(-n, 1 + c_{10} + c_{11} + n; c_{10} + 1; c_3 s) \tag{17}$$

where $c_{12} > 0$, $c_{13} > 0$ and $s \in [0, 1/c_3]$, $c_3 \neq 0$.

## 3. Arbitrary $l$-state solutions for the sDF oscillator potential

To study any quantum physical system characterized by the empirical exponential potential (1), we solve the following SE [1,2]

$$\left(\frac{P^2}{2\mu}+V(r)\right)\psi(r,\theta,\varphi)=E\psi(r,\theta,\varphi), \tag{18}$$

where $V(r)$ is taken as sDF oscillatory potential in Eq. (1b) and $\mu$ is the reduced mass. Inserting the wave functions $\psi(r,\theta,\varphi)=u(r)Y_{lm}(\theta,\varphi)$, with $u(r)=R(r)/r$ into Eq. (18), we obtain the radial SE for any $l$-state as

$$\left[\frac{d^2}{dr^2}+\frac{2\mu}{\hbar^2}\left(E_{nl}-D\left[\frac{b^2}{(e^{\alpha r}-1)^2}-\frac{2b}{e^{\alpha r}-1}\right]\right)-\frac{l(l+1)}{r^2}\right]R_{nl}(r)=0. \tag{19}$$

In Fig. 2, we plot effective sDF potential, i.e. $V_{\mathit{eff}}(r)=V_{sDF}(r)+l(l+1)/r^2$, with different values of centrifugal term for the $H_2$ diatomic molecule. Because of the centrifugal term, Eq. (19) can not be solved analytically for $l \neq 0$. Therefore, we attempt to use the following improved new approximation scheme to deal with this term near the minimum point $r=r_e$ [34,38]

$$\frac{1}{r^2}\approx\alpha^2\left[d_0+\frac{e^{-\alpha r}}{(1-e^{-\alpha r})^2}\right]$$

$$=\alpha^2\left[d_0+\frac{1}{(\alpha r)^2}-\frac{1}{12}+\frac{(\alpha r)^2}{240}-\frac{(\alpha r)^4}{6048}+\frac{(\alpha r)^6}{172800}+O\left((\alpha r)^8\right)\right]. \tag{20}$$

In the limiting case when $\alpha r \ll 1$, the value of the dimensionless constant $d_0=1/12$ and the screening parameter $\alpha$ has the unit of $1/r$. It is found that the approximation (20) surpasses the usual approximation [39,40] and reduces to $1/r^2$ once the parameter $\alpha$ goes to zero, that is,

$$\lim_{\alpha\to 0}\alpha^2\left(d_0+\frac{1}{e^{\alpha r}-1}+\frac{1}{(e^{\alpha r}-1)^2}\right)=\frac{1}{r^2}. \tag{21}$$

Now, substituting (21) into (19) and using the transformation of variables $s=e^{-\alpha r}$ which maps the half-line $(0,\infty)$ into the interval $[0,1]$, that maintains the finiteness of the transformed wave functions on the boundaries, we then obtain

$$\left\{\frac{d^2}{ds^2}+\frac{1-s}{s(1-s)}\frac{d}{ds}+\frac{1}{\alpha^2 s^2(1-s)^2}\right.$$

$$\left.\times\left[\varepsilon_{nl}(1-s)^2-db\left((2+b)s^2-2s\right)-\alpha^2 l(l+1)\left(d_0(1-s)^2+s\right)\right]\right\}R_{nl}(r)=0, \tag{22a}$$

$$\varepsilon_{nl}=\frac{2\mu E_{nl}}{\hbar^2} \text{ and } d=\frac{2\mu D}{\hbar^2}. \tag{22b}$$

Further, comparing Eq. (22) with Eq. (3), we obtain the following constants:

$$c_1 = 1, \qquad A = \frac{1}{\alpha^2}\left(db(2+b) - \varepsilon_{nl}\right) + l(l+1)d_0,$$

$$c_2 = 1, \qquad B = \frac{2}{\alpha^2}(db - \varepsilon_{nl}) + l(l+1)(2d_0 - 1),$$

$$c_3 = 1, \qquad C = -\frac{1}{\alpha^2}\varepsilon_{nl} + l(l+1)d_0. \qquad (23)$$

The remaining coefficients can be found via Eq. (6) which are displayed in table 1. By using Eq. (11) and constants in table 1, one can easily find the energy formula

$$E_{nl} = \frac{\hbar^2}{2\mu}l(l+1)\alpha^2 d_0 - \frac{\hbar^2\alpha^2}{2\mu}\left[\frac{\frac{\mu}{\hbar^2\alpha^2}b(2+b)D}{n+\delta_l} - \frac{n+\delta_l}{2}\right]^2, \qquad (24)$$

with

$$\delta_l = \frac{1}{2}\left(1 + \sqrt{(1+2l)^2 + \frac{8\mu}{\hbar^2\alpha^2}Db^2}\right) \geq 1. \qquad (25)$$

In Table 2, we present the potential parameters taken from Refs. [6,41,42] and used to generate the energy spectra for a few diatomic molecules $H_2$, $LiH$, $CO$ and $HCl$. Our results are given in table 3 together with the ones for the Morse potential [41]. It is worthy to note that $\hbar c = 1973.29\, eVA^0$ [4,11,41-43] and $1\, amu = 931.494028\, MeV/c^2$ [44] are used in the present calculations. Also, we obtained results numerically by using the amplitude phase (AP) method [45-48] to test the accuracy of our results. We noticed from Table 3 that the present approximation works well for the lowest states since this approximation is derived for the case $\alpha$ approaches 0 (cf. Eq. (21)). In Ref. [41], one of us used a parametric NU method derived for any exponential-type potential to obtain the bound state solutions of the spatially-dependent mass Schrödinger equation with the generalized $q$-deformed Morse potential for any rotational quantum number $l$. However, the results of the present work are found using a more generalized form of the exponential-type potential and also a different approximation formula.

Let us now turn to the calculation of the radial wave functions. Using Eq. (17) and constants in table 1, one finds

$$R_{nl}(r) = N_{nl} e^{-\eta\alpha r}(1-e^{-\alpha r})^{\delta_l} P_n^{(2\eta, 2\delta_l-1)}\left(1-2e^{-\alpha r}\right),$$

$$= N_{nl} \frac{(2\eta+1)_n}{n!} e^{-\eta\alpha r}(1-e^{-\alpha r})^{\delta_l} {}_2F_1\left(-n, n+2\eta+2\delta_l; 1+2\eta; e^{-\alpha r}\right), \qquad (26)$$

with

$$\eta = \frac{\frac{\mu}{\hbar^2 \alpha^2} b(2+b)D}{n+\delta_l} - \frac{n+\delta_l}{2} \qquad (27)$$

and $N_{nl}$ is the normalization constant. Here $P_n^{(a,b)}(x)$ is the Jacobi polynomial and ${}_2F_1(-n,\beta;\gamma;z)$ is the hypergeometric function. The normalization constant $N_{nl}$ can be found in closed form by using the normalization condition: $\int_0^\infty |R_{nl}(r)|^2 dr = \int_0^1 |R_{nl}(s)|^2 \frac{ds}{\alpha s} = 1$, where $s = e^{-\alpha r}$, and obtain

$$|N_{nl}|^2 \int_0^1 s^{2\eta-1}(1-s)^{2\delta_l}\left[{}_2F_1(-n,n+2\eta+2\delta_l;2\eta+1;s)\right]^2 ds = \alpha\left(\frac{n!\Gamma(2\eta+1)}{\Gamma(2\eta+n+1)}\right)^2. \qquad (28)$$

With the help of the formula [49,50]

$$\int_0^1 s^{2a-1}(1-s)^{2(b+1)}\left[{}_2F_1(-n, n+2(a+b+1); 2a+1; s)\right]^2 ds$$

$$= \frac{(n+b+1)n!\Gamma(n+2b+2)\Gamma(2a)\Gamma(2a+1)}{(n+a+b+1)\Gamma(n+2a+1)\Gamma(n+2(a+b+1))}, \quad a > -1/2,\ b > -3/2, \qquad (29)$$

and setting $a = \eta$ and $b = \delta_l - 1$, we then calculate the normalization constant from (28) as

$$N_{nl} = \sqrt{\frac{2\eta\alpha n!(n+\eta+\delta_l)\Gamma(n+2(\eta+\delta_l))}{(n+\delta_l)\Gamma(n+2\eta+1)\Gamma(n+2\delta_l)}}. \qquad (30)$$

For the ground state, $n = 0$, the normalization factor is

$$N_{0l} = \sqrt{\frac{\alpha(\eta+\delta_l)}{\delta_l B(2\eta, 2\delta_l)}}, \quad B(2\eta, 2\delta_l) = \frac{\Gamma(2\eta+1)\Gamma(2\delta_l)}{2\eta\Gamma(2\eta+2\delta_l)}. \qquad (31)$$

## 4. Conclusion

In this work, we have obtained the bound state solutions of the radial SE with the shifted Deng-Fan oscillator potential in the framework of the parametric NU method. The analytical expressions for the energy eigenvalues and the corresponding wave functions are obtained in closed form. In Table 3, we present the numerical ro-vibrational energy states of a few diatomic molecules with various arbitrary values of rotational and vibrational quantum numbers. It is worthy to note that in the examples considered the Deng-Fan oscillator potential does not predict observed energy levels significantly better than the Morse potential despite its correct asymptotic behavior. The present numerical results are in good agreement with those ones obtained previously by using the Morse oscillator potential.

**Acknowledgments**

We thank the kind referees for the positive suggestions and critics which have greatly improved the present paper. S.M. Ikhdair acknowledges the partial support of the Scientific and Technological Research Council of Turkey. M. Hamzavi thanks his host Institution, KTH-Mechanics, Royal Institute of technology, S-100 44 Stockholm, Sweden.

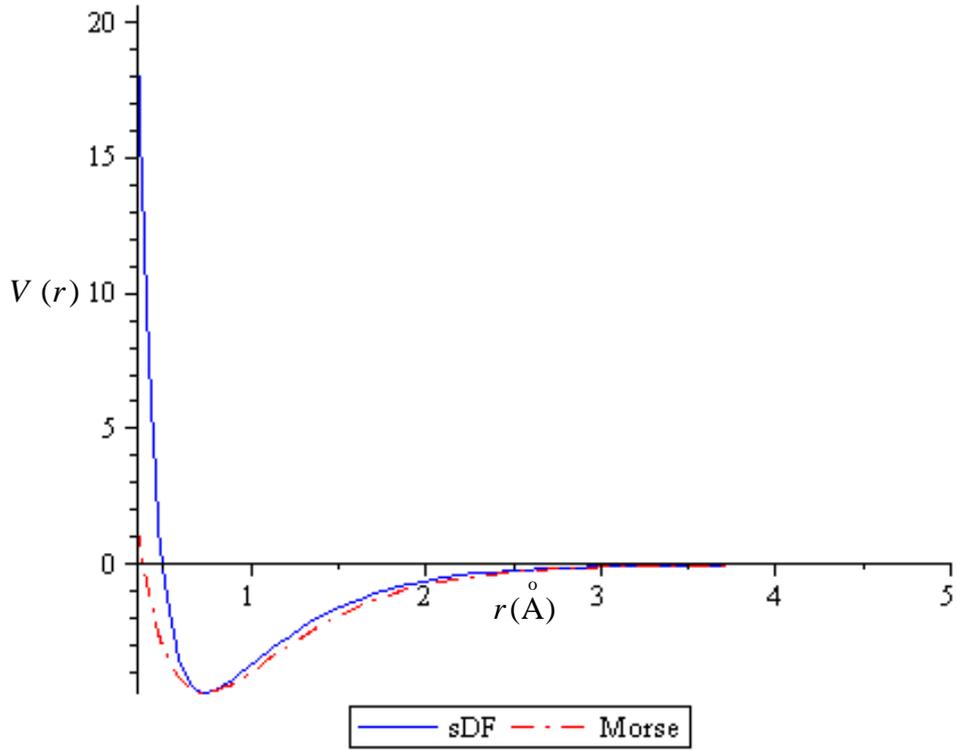

**Figure 1.** Shape of the sDF and Morse oscillator potentials for $H_2$ diatomic molecule.

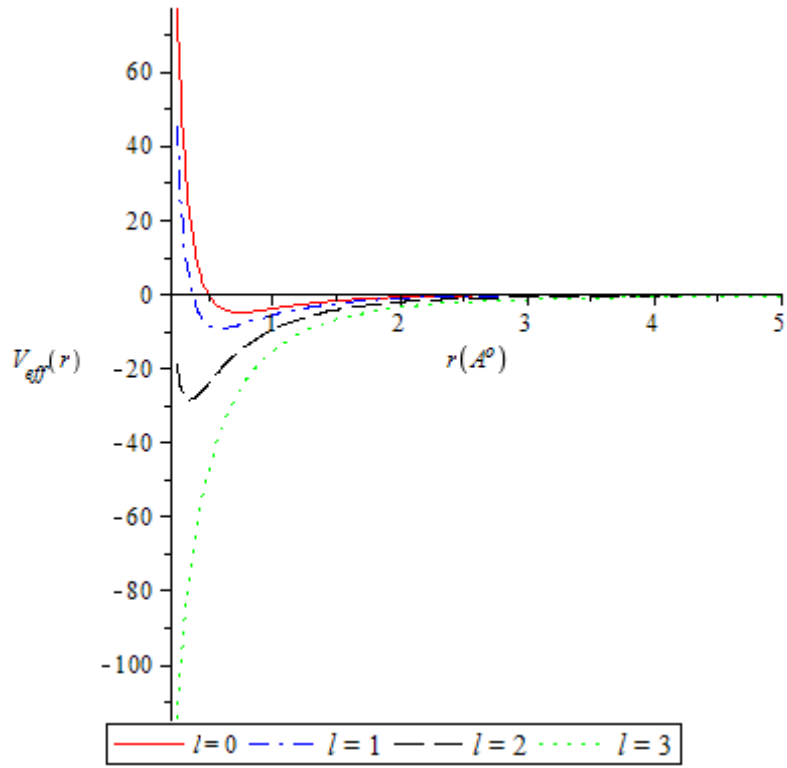

**Figure 2.** Shape of the sDF together with different values of centrifugal term for $H_2$ diatomic molecule.

**Table 1.** The specific values for the parametric constants.

| Constant | Analytical value |
|---|---|
| $c_4$ | $0$ |
| $c_5$ | $-\dfrac{1}{2}$ |
| $c_6$ | $A+\dfrac{1}{4}$ |
| $c_7$ | $-B$ |
| $c_8$ | $C$ |
| $c_9$ | $\dfrac{1}{4}\left[4(A-B+C)+1\right]$ |
| $c_{10}$ | $2\sqrt{C}$ |
| $c_{11}$ | $\sqrt{4(A-B+C)+1}$ |
| $c_{12}$ | $\sqrt{C}$ |
| $c_{13}$ | $\dfrac{1}{2}\left[1+\sqrt{4(A-B+C)+1}\right]$ |

**Table 2.** The potential model parameters for some diatomic molecules.

| Molecule | $\mu(amu)$ | $a(A^{°-1})$ | $r_e(A^°)$ | $D(cm^{-1})$ |
|---|---|---|---|---|
| $H_2$ [41] | 0.50391 | 1.9426 | 0.7416 | 38266 |
| $LiH$ [6] | 0.8801221 | 1.1280 | 1.5956 | 20287 |
| $CO$ [6] | 6.8606719 | 2.2994 | 1.1283 | 90540 |
| $HCl$ [6] | 0.9801045 | 1.8677 | 1.2746 | 37255 |

**Table 3.** The energy levels for a few diatomic molecules obtained from the sDF and Morse oscillator potentials.

| $n$ | $l$ | $-E_{nl}$ (eV) NU | $-E_{nl}$ (eV) AP | $-E_{nl}$ (eV) [41] |
|---|---|---|---|---|
| $H_2$ | | | | |
| 0 | 0 | 4.39444 | 4.39444 | 4.47601 |
|   | 5 | 4.17644 | 4.18054 | 4.25880 |
|   | 10 | 3.62165 | 3.63782 | 3.72194 |
| 5 | 0 | 1.75835 | 1.75835 | 2.22052 |
|   | 5 | 1.61731 | 1.62548 | 2.04355 |
|   | 10 | 1.26034 | 1.29257 | 1.60391 |
| 7 | 0 | 1.07756 | 1.07756 | 1.53744 |
|   | 5 | 0.96174 | 0.97232 | 1.37656 |
|   | 10 | 0.66976 | 0.71172 | 0.97581 |
| LiH | | | | |
| 0 | 0 | 2.41195 | 2.41195 | 2.42886 |
|   | 5 | 2.38348 | 2.38458 | 2.40133 |
|   | 10 | 2.30815 | 2.31229 | 2.32884 |
| 5 | 0 | 1.51628 | 1.51628 | 1.64771 |
|   | 5 | 1.49278 | 1.49429 | 1.62377 |
|   | 10 | 1.43062 | 1.43627 | 1.56074 |
| 7 | 0 | 1.22340 | 1.22340 | 1.37756 |
|   | 5 | 1.20173 | 1.20344 | 1.35505 |
|   | 10 | 1.14444 | 1.15083 | 1.29580 |
| CO | | | | |
| 0 | 0 | 11.08068 | 11.08068 | 11.0915 |
|   | 5 | 11.07247 | 11.07354 | 11.0844 |
|   | 10 | 11.05057 | 11.05449 | 11.0653 |
| 5 | 0 | 9.68809 | 9.68809 | 9.79518 |
|   | 5 | 9.68017 | 9.68130 | 9.78833 |
|   | 10 | 9.65905 | 9.66321 | 9.77009 |
| 7 | 0 | 9.15911 | 9.15911 | 9.29918 |
|   | 5 | 9.15131 | 9.15247 | 9.29246 |
|   | 10 | 9.13050 | 9.13476 | 9.27455 |
| HCl | | | | |
| 0 | 0 | 4.41705 | 4.41705 | 4.43556 |
|   | 5 | 4.37403 | 4.37843 | 4.39682 |
|   | 10 | 4.25973 | 4.27591 | 4.29408 |
| 5 | 0 | 2.66574 | 2.66574 | 2.80506 |
|   | 5 | 2.62859 | 2.63411 | 2.77209 |
|   | 10 | 2.52989 | 2.55027 | 2.68471 |
| 7 | 0 | 2.09652 | 2.09652 | 2.25701 |
|   | 5 | 2.06161 | 2.06768 | 2.22634 |
|   | 10 | 1.96888 | 1.99127 | 2.14511 |